\documentclass[letterpaper, 10 pt, conference]{ieeeconf}  

\IEEEoverridecommandlockouts                              
\usepackage{graphicx,scalefnt,csquotes}
\usepackage{amsmath,graphicx,scalefnt}
\title{\LARGE \bf
Effects of Feedback Latency on P300-based Brain-computer Interface}

\author{Mahnaz Arvaneh$^{1}$, Tomas E. Ward $^{2}$, and Ian H. Robertson$^{1}$ 
\thanks{*This work was supported by Science Foundation Ireland (SFI) under
Grant No. 12/RC/2289.}
\thanks{$^{1}$ M. Arvaneh and I. H. Robertson are with Trinity College Institute of Neuroscience, and Insight Centre for Data Analytics, Dublin, Ireland (emails:arvanehm,iroberts@tcd.ie)}%
\thanks{$^{2}$ T. E. Ward is
with Dept. of Electronic Engineering, National University of
Ireland, Maynooth, Ireland. (email:
tomas.ward@eeng.nuim.ie).}%
}

\begin{document}

\maketitle
\thispagestyle{empty}
\pagestyle{empty}

\begin{abstract}
Feedback has been shown to affect performance when using a
Brain-Computer Interface (BCI) based on sensorimotor rhythms. In
contrast, little is known about the influence of feedback on
P300-based BCIs. There is still an open question whether feedback
affects the regulation of P300 and consequently the operation of
P300-based BCIs. In this paper, for the first time, the influence of
feedback on the P300-based BCI speller task is systematically
assessed. For this purpose, 24 healthy participants performed the
classic P300-based BCI speller task, while only half of them
received feedback. Importantly, the number of flashes per letter was
reduced on a regular basis in order to increase the frequency of
providing feedback. Experimental results showed that feedback could
significantly improve the P300-based BCI speller performance, if it
was provided in short time intervals (e.g. in sequences as short as
4 to 6 flashes per row/column). Moreover, our offline analysis
showed that providing feedback remarkably enhanced the relevant ERP
patterns and attenuated the irrelevant ERP patterns, such that the
discrimination between target and non-target EEG trials increased.
\end{abstract}

\section{INTRODUCTION}
A brain-computer interface (BCI) provides a direct communication
pathway between a human brain and an external device \cite{cit:0}.
Using appropriate sensors and data processing algorithms, a BCI maps
patterns of brain activity associated with a volitional thought onto
commands suitable for controlling a device \cite{cit:1}. Such
technology can be potentially used as an assistive device, a
rehabilitation tool or a brain training protocol \cite{cit:2}. In
most BCI systems, brain signals are measured by electroencephalogram
(EEG), due to its low cost and high temporal resolution
\cite{cit:3}. Currently, majority of the EEG-based BCI systems are
working based on either P300 \cite{cit:4} or sensorimotor rhythms
\cite{cit:3}.


A BCI is a closed-loop system relying on mutual learning efforts
between the user who learns to generate robust EEG patterns, and the
classification system that is trained to accurately identify the EEG
patterns. The mutual learning between the user and the classifier is
crucial for having an accurate and robust BCI system. However, most
of the BCI studies have only focused on training of the
classification system, and neglected the user learning part.

Typically, the user learning is mediated by feedback provided from
the classifier. Feedback is known to significantly increase the
motivation of learning \cite{cit:5}. Although, several studies
showed the effectiveness of feedback on high quality learning
\cite{cit:6}, a poorly designed feedback mechanism may deteriorate
the performance \cite{cit:5}. There are some interesting studies
considering effects of different types of feedback on motor
imagery-based BCIs. For example, McFarland et al. suggested that
feedback facilitates initial learning of the BCI skill \cite{cit:7}.
Neuper et al. showed that continuous feedback is more efficient than
delayed discrete feedback \cite{cit:8}. Some authors explored
multidimensional feedback (e.g. 3D or Virtual Reality feedback)
\cite{cit:9}, whereas some studies investigated the effects of
biased feedback, negative feedback, and positive feedback on motor
imagery-based BCIs \cite{cit:10,cit:11}.

Despite all these studies, little is known about the influence of
feedback on P300-based BCIs. There is still an open question whether
feedback affects the regulation of P300 and consequently the
operation of P300-based BCIs. It seems logical to expect that
feedback could positively influence the performance of P300-based
BCIs, since, in addition to learning effects, feedback increases
motivation, and motivation modulates the P300 amplitude during BCI
use as shown in \cite{cit:12}. Nevertheless, in the study conducted
by McFarland et al. providing feedback did not affect the P300-based
BCI results \cite{cit:4}. There might be a number of reasons leading
to this observation. Importantly, there is a substantial delay
between the stimuli and the feedback, since the feedback is given
after averaging a relatively high number of trials. Thus, the user
cannot be certain in which trials he/she behaved incorrectly.
Moreover, due to averaging over a large number of trials, the user
usually achieves a high accuracy. It means negative feedback is
rarely given. Alternatively, the processes involved in the
generation of P300 may not be readily influenced by feedback.

This paper investigates the above-mentioned possibilities. In this
paper, as the first study, we systematically assesses the influence
of feedback on the P300-based BCI speller system. For this purpose,
24 healthy participants performed the classic P300-based BCI speller
task, while only half of them received feedback. Importantly, to
provide more frequent feedback the number of trials for averaging
was reduced on a regular basis. Providing feedback based on fewer
number of trials also increased the probability of having negative
feedback. In addition to the online accuracy, the effects of
feedback on the EEG patterns are also explored by considering the
power of the target trials versus the power of the non-target
trials.
\section{MATERIALS AND METHODS}
\subsection{Participants}

In total, 24 healthy young adults aged 18-39 years were participated
in this study.  The participants had no history of neurological
illness. They gave informed consent to the study which had been
reviewed and approved by the ethical review board of the School of
Psychology, Trinity College Dublin, in accordance with the
Declaration of Helsinki.

Participants were randomly assigned to two groups, labeled as
``Feedback" and ``No-feedback". The Feedback group consisted of 5
females and 7 males with a mean age of 27.25 (SD$\pm6.00$). The
No-feedback group consisted of 7 females and 5 males with a mean age
of 24.83 (SD$\pm3.09$). 22 of the entire 24 participants had no
previous experience of performing BCI sessions. There were no
significant differences between the two groups according to the
demographic variables (age: $\mathrm{t}(22)\!\!=\!1.22,
\mathrm{p}\!\!=\!0.24$; gender: $\chi^{2}(1)\!\!=\!0.67,
\mathrm{p}\!\!=\!0.48$; and BCI naivety: $\chi^{2}(1)\!\!=\!2.18,
\mathrm{p}\!\!=\!0.14)$.

\subsection{EEG data acquisition}
EEG was acquired using a Biosemi ActiveTwo system from 8 electrodes
located at positions Fz, C3, Cz, C4, P3, Pz, P4, and Oz following
the International 10-20 system. Impedances were kept below $5
\mathrm{k}\Omega$, and the sampling rate was 512 Hz. The P300-based
speller task was designed using the BCI2000 software
\cite{cit:10_1}. In offline analysis, the continuous EEG data were
filtered with a zero-phase low-pass 35 Hz Butterworth filter and a
zero-phase high-pass 0.5 Hz Butterworth filter. The EEG data were
segmented and baseline corrected relative to the interval -150 to 0
ms before the onset of the stimuli. Segments with amplitudes
exceeding $+75\mu V$, or voltage steps of more than 150$\mu V$
within a window of 200 ms were rejected from further analysis.

\subsection{P300-based BCI speller task}
\begin{figure}[htb]
\begin{minipage}[b]{1.0\linewidth}
  \centering
  \centerline{\includegraphics[width=7cm, height=5cm]{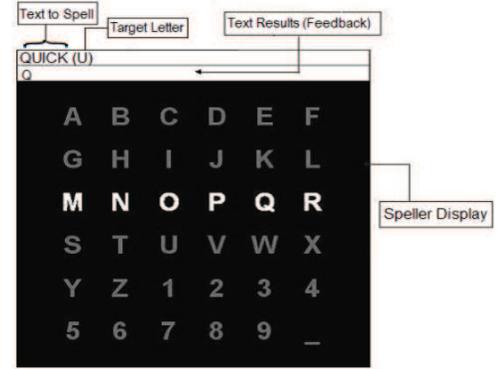}}
\end{minipage}
\vspace{-0.6cm}
 \caption{The interface of the P300-based BCI speller task}
\label{fig:speller} \vspace{-0.5 cm}
\end{figure}

The P300-based speller task has been widely used in BCI community as
an assistive device for communication \cite{cit:4}. Fig.
\ref{fig:speller} presents the interface of the P300-based speller
task used in this study. A $6\!\times\!6$ matrix, containing the
letters of the alphabet and other symbols, was displayed on a
computer screen, while EEG was recorded. The text-to-spell was
presented above the matrix. Directly next to the text-to-spell, the
target letter-to-spell was displayed in the parenthesis. The rows
and the columns of the matrix were flashed/intensified in a random
order. Each flash lasted 55 ms followed by an inter stimulus
interval of 117 ms. The participants were instructed to concentrate
on the target letter and silently count how often it flashed.
Basically, flashes of the row and the column containing the target
letter (i.e. target stimuli) evoke P300 on the EEG signals, whereas
flashes of the other rows and columns (i.e. non-target stimuli)
correspond to neutral EEG signals. Thus, the target letter can be
inferred by a classification algorithm that searches for the row and
the column which evoked the largest P300 responses.

%
After each sequence, the matrix stopped flashing for 6 seconds.
During this interval, the next target letter was displayed in the
parentheses. Thus, the participant was allowed sufficient time to
locate this new target in the matrix. This process was repeated over
the entire target word.

\subsection{Procedure}
\begin{figure}[htb]
\begin{minipage}[b]{1.0\linewidth}
  \centering
  \centerline{\includegraphics[width=7cm, height=5cm]{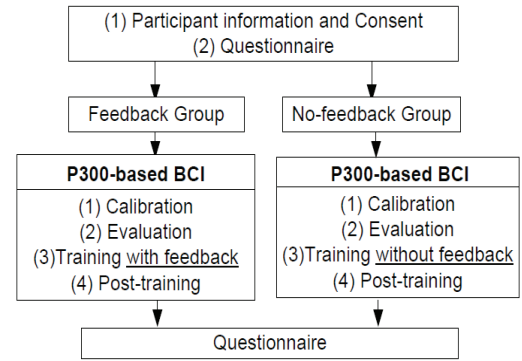}}
\end{minipage}
\vspace{-0.6 cm}
 \caption{A schematic illustration of the procedure in the present study}
\label{fig:schematic} \vspace{-0.3 cm}
\end{figure}

The procedure of the present study is illustrated in Fig.
\ref{fig:schematic}. Participants engaged in a single session of
around one hour duration including set up time. The data recording
and task performance took place in a dark sound-attenuated closed
room.

\subsubsection{Questionnaire}
At the beginning and at the end of the session, the participants
filled out a simple questionnaire which was used to record
self-reported levels of alertness, boredom, tiredness of the mind,
and tiredness of the eyes on a 10-point Likert scale.

\subsubsection{Calibration} In this stage, the participants completed two
runs, in which the words ``the" and ``quick" were respectively
spelled without providing feedback. The sequence of attending each
letter consisted of 12 flashes per row/column. The EEG data
collected from this stage were used to calibrate a subject-specific
model identifying attended letters.

After completing the two calibration runs, the EEG responses to each
row/column was obtained by averaging over the 12 corresponding EEG
trials. Thereafter, 800 ms segments starting immediately after the
onset of the stimuli were extracted from the EEG responses. The
segments were decimated to 20 Hz. Then, the resulting data arrays
were concatenated to form the feature vector. The dimension of the
feature vector was $N_e\!\times\!N_t$, where $N_e$ denotes the
number of electrodes and $N_t$ denotes the number of temporal
samples in an EEG response. Finally, the extracted features were
used to train a linear discriminant analysis (LDA) classifier which
was used to discriminate between target and non-target trials. It
should be noted that the specific P300 paradigm presented here has
been demonstrated to yield reliable
performance in several studies \cite{cit:12,cit:13}. 

\subsubsection{Evaluation} In this stage, the participants were asked to spell the word
``dog" without receiving feedback. The EEG data collected from this
stage were used to evaluate the model calibrated in the calibration
stage. If at least two of the three letters of ``dog" were
identified correctly, the participant was ready to move to the
training stage. Otherwise, the participant was removed from the
study. It is noted that in this study all the participants were able
to achieve a calibration model with satisfactory performance.

\subsubsection{Training} In this stage, the participants were asked to spell the word
``beautiful" in 4 runs. The number of flashes per row/colum was set
to 10, 8, 6, and 4 in the first, the second, the third and the
fourth run respectively.

For the Feedback group, feedback was provided at the end of each
sequence of flashes through presentation of the attended letter
target as determined by the output of the classifier. By reducing
the number of flashes participants received feedback more
frequently. Thus, this presents a means by which we can investigate
the effects of feedback on BCI performance. Significantly reducing
the number of flashes can increase the probability of receiving
negative feedback as the decision of the classifier becomes more
dependent on each individual P300 response. The participants in the
No-feedback group underwent entirely similar process, but without
receiving any feedback. They were not informed about the output of
the classifier neither on the screen nor orally. This is not a
situation which one encounters with a P300 BCI in normal operation
but it is necessary here in order to isolate the effects of feedback
explicitly.

\subsubsection{Post-training} In this stage, the Feedback and the No-feedback
groups both spelled the word ``dance", without receiving feedback.
Similar to the calibration stage, in this stage the number of
flashes per row/colum was set to 12.

\section{RESULTS AND DISCUSSION }
\subsection{Questionnaire}
Analysis of the pre-scores and the post-scores suggested that both
the groups (Feedback and No-feedback) were similar in terms of
alertness, boredom, tiredness of the mind, and tiredness of the
eyes. Furthermore, the within-group comparisons revealed no
significant changes on any of the variables prior and following the
test.

\subsection{Effects of feedback on classification accuracy}
\begin{figure}[htb]
\begin{minipage}[b]{1.0\linewidth}
  \centering
  \centerline{\includegraphics[width=6cm, height=3cm]{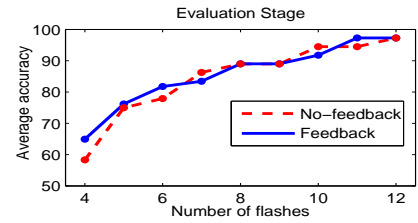}}
\end{minipage}
\vspace{-0.7 cm}
 \caption{Average classification accuracy as a function of number of flashes per row/column in the evaluation stage (no feedback was provided).}
\label{fig:Eval} \vspace{-0.2 cm}
\end{figure}

Results obtained from the evaluation stage showed that both groups
were very successful in performing the task, since they achieved
similar average classification accuracies of $97.25\pm9.6$ on
spelling the word ``fox". Obtained from offline analysis, Fig.
\ref{fig:Eval} shows the average classification accuracies of the
evaluation stage as a function of the number of flashes per
column/row. A repeated ANOVA revealed a significant main effect for
the number of flashes ($\mathrm{F}(11,242)\!\!=\!39.72,
\mathrm{p}\!<\!0.001$). Neither the group
($\mathrm{F}(1,22)\!\!=\!0.03, \mathrm{p}\!\!=\!0.86$) nor the
interaction between the group and the number of flashes
($\mathrm{F}(1,11)\!\!=\!0.18, \mathrm{p}\!\!=\!0.99$) had a
significant effect. The statistical analysis confirms that the
Feedback and the No-feedback groups were very similar in performing
the P300-based speller task in the evaluation stage.

\begin{figure}[htb]
\begin{minipage}[b]{1.0\linewidth}
  \centering
  \centerline{\includegraphics[width=8cm, height=4.5cm]{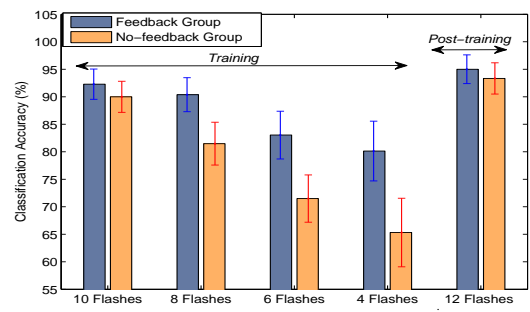}}
\end{minipage}
\vspace{-0.7 cm}
 \caption{Average classification accuracies obtained from Feedback and No-feedback groups over the training and post-training stages}
\label{fig:acc} \vspace{-0.2 cm}
\end{figure}

In the training stage, the word ``beautiful" was spelled 4 times
using different number of flashes per column/row. Fig. \ref{fig:acc}
presents the average classification results obtained from the
training stage as well as the post-training stage where the word
``dance" was spelled without providing feedback to both of the
groups. As Fig. \ref{fig:acc} shows the feedback group outperformed
the No-feedback group in terms of the classification accuracy.
Interestingly, a smaller number of flashes yielded a larger
difference between the performance of these two groups. A repeated
ANOVA revealed significant main effects for the number of flashes
($\mathrm{F}(3,66)\!\!=\!15.74, \mathrm{p}\!<\!0.001$) and the group
($\mathrm{F}(1,22)\!\!=\!4.74, \mathrm{p}\!\!=\!0.04$). However, the
interaction between the group and the number of flashes
($\mathrm{F}(1,11)\!\!=\!2.08, \mathrm{p}\!\!=\!0.11$) was not
significant. Since the Feedback and the No-feedback groups performed
very similarly during the evaluation stage, the significant
difference between these two groups during the training stage is
most likely due to feedback provided to the Feedback group.
Exploratory analysis using independent t-tests indicated that the
Feedback group significantly performed better than the No-feedback
group when the number of flashes was 4 ($\mathrm{t}(22)\!\!=\!2.1,
\mathrm{p}\!\!=\!0.04$). The superior performance of the Feedback
group tended to be significant when the number of flashes was 6
($\mathrm{t}(22)\!\!=\!1.96, \mathrm{p}\!\!=\!0.06$). However, no
significant differences in performance was found for greater flash
sequences ($\mathrm{p}\!\!=\!0.10$ and $\mathrm{p}\!\!=\!0.53$ for 8
and 10 flashes respectively). These results suggest that the effect
of feedback is more pronounced when it is given in shorter time
intervals (i.e. Feedback is more frequent).

Importantly, in the post-training stage where both groups did not
receive any types of feedback, again the classification results were
similar ($\mathrm{F}(1,22)\!\!=\!0.71, \mathrm{p}\!\!=\!0.41$).
Offline analysis of the classification results as a function of the
number of flashes per column/row revealed a significant main effect
for the number of flashes as expected
($\mathrm{F}(11,242)\!\!=\!38.96, \mathrm{p}\!<\!0.001$). Neither
the group ($\mathrm{F}(1,22)\!\!=\!0.03, \mathrm{p}\!\!=\!0.95$) nor
the interaction between the group and the number of flashes
($\mathrm{F}(1,11)\!\!=\!1.95, \mathrm{p}\!\!=\!0.13$) had a
significant effect.

\subsection{Effects of feedback on EEG patterns}
In addition to the classification accuracy, we conducted an offline
re-analysis of the data to better understand the effects of feedback
on EEG patterns. We speculated that by receiving feedback, subjects
may consciously or unconsciously modify patterns of his/her brain
activity throughout the experiment. To seek changes in EEG patterns,
we defined a new criterion, called as signal to noise ratio (SNR).
SNR was calculated by the average power of the target trials (i.e.
signal) divided by the average power of the non-target trials (i.e.
noise) over the centroparietal electrodes (namely C3, Cz, C4, P3,
Pz, P4). Since N200 and P300 are both contributing in the
classification of P300-based BCIs \cite{cit:13}, 150 ms to 550 ms
after the onset of the stimuli were used for calculating the
energies.

\begin{figure}[htb]
\begin{minipage}[b]{1.0\linewidth}
  \centering
  \centerline{\includegraphics[width=8.5cm, height=3.5cm]{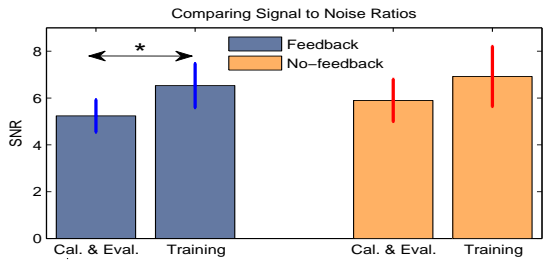}}
\end{minipage}
\vspace{-0.7 cm}
 \caption{Comparing the signal to noise ratios (SNR) between the calibration and evaluation stages vs. the training stage for Feedback and No-feedback groups. Asterisk indicates a tending to be significant
difference ($p=0.07$).} \label{fig:SNR} \vspace{-0.2cm}
\end{figure}

Fig. \ref{fig:SNR} compares the average SNR value obtained from the
calibration and evaluation stages with the average SNR value
obtained from the training stage. Indeed, neither the Feedback group
nor the No-feedback group received feedback in the calibration and
evaluation stages, whereas the feedback was provided to the Feedback
group during the training stage. There was no significant difference
between the average SNR values of the two groups during the
calibration and the evaluation stages ($\mathrm{F}(1,22)=0.33,
\mathrm{p}=0.57$). As Fig. \ref{fig:SNR} shows, on average both
groups presented improvements on SNR, when they transferred to the
training stage. This improvement might be due to learning to better
attend the task as time passes regardless of providing feedback.
Besides, reducing the number of flashes in the training stage leads
to shorter sequences of flashes for spelling each word. Attending to
shorter sequences might be less distracting than attending to longer
sequences. Interestingly, the paired t-test revealed that the
improvement in the SNR values of the feedback group tends to be
significant ($\emph{t}(11)\!\!=\!-1.99, \emph{p}\!\!=\!0.07$.
However, the improvement in the SNR values of the No-feedback group
was not significant $\emph{t}(11)\!\!=\!-1.02, \emph{p}\!\!=\!0.33$.
Thus, we can conclude that there is an association between feedback
and the improvement in SNR.

\section{CONCLUSIONS}
With this study, we showed that feedback can positively influence
the performance of P300-based BCIs, if it is provided more often
than what most P300-based BCI systems currently provide. Our
experiments revealed that when feedback was given after sequences of
4 flashes per row/column, the Feedback group significantly
outperformed the No-feedback group ($\mathrm{p}\!\!=\!0.04$),whereas
the superior performance of the Feedback group tended to be
significant when the number of flashes was 6
($\mathrm{p}\!\!=\!0.06$). However, the effect of feedback was not
significant as the number of flashes increased. These findings are
unsurprising, because when the feedback is provided after a large
number of flashes the user cannot be certain in which trials he/she
behaved incorrectly. Moreover, the user may not even get aware of
failure in attention in some trials, since it can be compensated by
averaging over many other trials. Reducing the number of flashes
mitigates this problem by directing the user to better attend the
task. Our study on the EEG patterns also showed that providing
feedback remarkably contributed in improving SNR.


\end{document}